\documentclass[conference]{IEEEtran}
\IEEEoverridecommandlockouts
\usepackage{cite}
\usepackage{amsmath,amsthm,amssymb,amsfonts}

\usepackage{graphicx}
\usepackage{textcomp}
\usepackage{xcolor}
\def\BibTeX{{\rm B\kern-.05em{\sc i\kern-.025em b}\kern-.08em
    T\kern-.1667em\lower.7ex\hbox{E}\kern-.125emX}}
\usepackage{multirow}
\usepackage{tabularx}
\usepackage{longtable}
\usepackage{siunitx}
\usepackage{dblfloatfix}
\usepackage{float}
\usepackage{xcolor}
\usepackage[font=footnotesize]{caption}

\DeclareCaptionLabelSeparator{periodspace}{.\quad}
\usepackage{cite}
\usepackage{multirow}
\usepackage{hyperref}
\usepackage{graphicx}
\usepackage{amsmath}
\usepackage{multicol}
\usepackage{mathtools}
\usepackage{tabularx}
\usepackage{longtable}
\usepackage{mdwlist}

\usepackage{algpseudocode}
\usepackage{balance}
\usepackage[square,numbers]{natbib}
\usepackage{booktabs}
\usepackage{hyperref}
\usepackage{url}

\begin{document}
\captionsetup[figure]{labelsep=colon,name={Fig.}}

\title{Analyzing Distribution Transformer Degradation with Increased Power Electronic Loads\\
% {\footnotesize \textsuperscript{*}Note: Sub-titles are not captured in Xplore and
% should not be used}
\thanks{This work was supported by the Sensors and Data Analytics Program of the U.S. Department of Energy Office of Electricity, under Contract No. DE-AC05-76RL01830.}
}

\author{\IEEEauthorblockN{Bhaskar Mitra, Ankit Singhal, Soumya Kundu and James P. Ogle}
\IEEEauthorblockA{\textit{Electricity Infrastructure and Buildings Division} \\
Pacific Northwest National Laboratory\\
Richland, WA 99354, USA \\
Email: \{bhaskar.mitra, ankit.singhal, soumya.kundu, james.ogle\}@pnnl.gov}}

\maketitle

\begin{abstract}
The influx of non-linear power electronic loads into the distribution network has the potential to disrupt the existing distribution transformer operations. They were not designed to mediate the excessive heating losses generated from the harmonics. To have a good  understanding of  current standing challenges, a knowledge of the generation and load mix as well as the current harmonic estimations are essential for designing transformers and evaluating their performance. In this paper, we investigate a mixture of essential power electronic loads for a household designed in PSCAD/EMTdc and their potential impacts on transformer eddy current losses and derating using harmonic analysis. The various scenarios have been studied with increasing PV penetrations. The peak load conditions are chosen for each scenario to perform a transformer derating analysis. Our findings reveal that in the presence of high power electronic loads (especially third harmonics), along with increasing PV generation may worsen transformer degradation. However, with a low amount of power electronic loads, additional PV generation helps to reduce the harmonic content in the current and improve transformer performance. 
\end{abstract}


\section{Introduction}
Power electronic loads have found a wider application in power system networks especially after their advancement in the late 1900s. Many loads require essential power electronic converters for stage conversion. Some common examples of power electronic loads include uninterrupted power supply (UPS) devices, personal computers, laptops, electric vehicle chargers, etc. These non-linear loads contribute to non-linear sinusoidal currents. The non-sinusoidal currents, when passing through network impedance, create a non-sinusoidal voltage drop \cite{grady2012understanding}. The non-sinusoidal voltage and current components are integer multiples of the fundamental component called \textit{``harmonics"}. The deterioration of the supply voltage creates stress on the electrical equipment and can potentially damage it, resulting in increased operating costs and downtime \cite{Fuchs2008}.

Increased voltage and current harmonics are found to have a direct relationship to premature aging and degradation of transformers. Initial transformer designs were made considering conventional load models, i.e., Constant Impedance (Z), Constant Current (I), and Constant Power (P) or \textit{``ZIP"} models, that would operate at fundamental 60Hz or 50Hz frequency \cite{McLorn2017,bokhari2013experimental}. Under the increased penetration of non-linear loads, the design of power transformers needs to be reassessed to ensure proper and safe operation. Increased non-linear loads increase the transformer losses due to overheating of the core, creating a larger derating factor \cite{Lavers1999,Masoum2008}.

The problem of harmonics is more evident with customers on the low-voltage end. The common household equipment includes but not limited to desktops, laptops, LED lamps, variable speed drives, solar panels, etc. To compound the challenges, as more and more electric vehicles come to the market, they rely majorly on at-home charging that produces a large fraction of non-linear voltage and current. Certain power electronic devices like VFD's contribute more $3^{rd}$ harmonics, if they are not properly compensated it would lead to additional losses and loss-of-life for the transformer. The addition of harmonics has an effect on transformer protection as well. Addition of the $5^{th}$ harmonic needs to be compensated below a certain threshold before the protection relays can be engaged. The distorted harmonic waveforms results in loss of essential information for protection, this might result in the protection devices operating slower \cite{jain2011harmonics}.

Currently, there is a gap in high-fidelity load models that can capture the typical characteristics of non-linear models, i.e., the cross-coupling effect of voltages and current. A harmonic-rich current/voltage dataset is essential to understand the effect on transformer losses, heating, etc. \textit{``ZIP"} based harmonic load models suffer from a lack of enhanced harmonic spectrum that can be observed through the operation of various non-linear devices \cite{collin_component-based_2010}. Real-world field data is not publicly available to perform such analysis. Methods relying on laboratory setups to develop such datasets fail to capture the effect on other nonlinear load currents.

Therefore, in this paper, (i) we perform an analysis of residential transformer heating and losses encountered due to the presence of non-linear power electronic residential loads using PSCAD/EMTdc. Detailed power electronic models are developed to create harmonic rich datasets to entail their effect on transformer operation; (ii) Different loading scenarios on the transformer are assessed with increasing PV penetration to understand its effect on THD(\%), eddy current losses, and the subsequent impact on transformer derating.

The rest of the paper has been organized in the following way. Section \ref{Modelling} discusses the modeling approach for the study. Section \ref{sec:trans} discusses the calculation of eddy current losses due to nonlinear harmonic current in transformers. Section \ref{Results} discusses the results, and Section \ref{Conclusion} concludes the paper with major findings and proposed future enhancements.

\section{Modeling} \label{Modelling}
\subsection{System Description}
\begin{figure}
    \centering
    \includegraphics[width=0.48\textwidth]{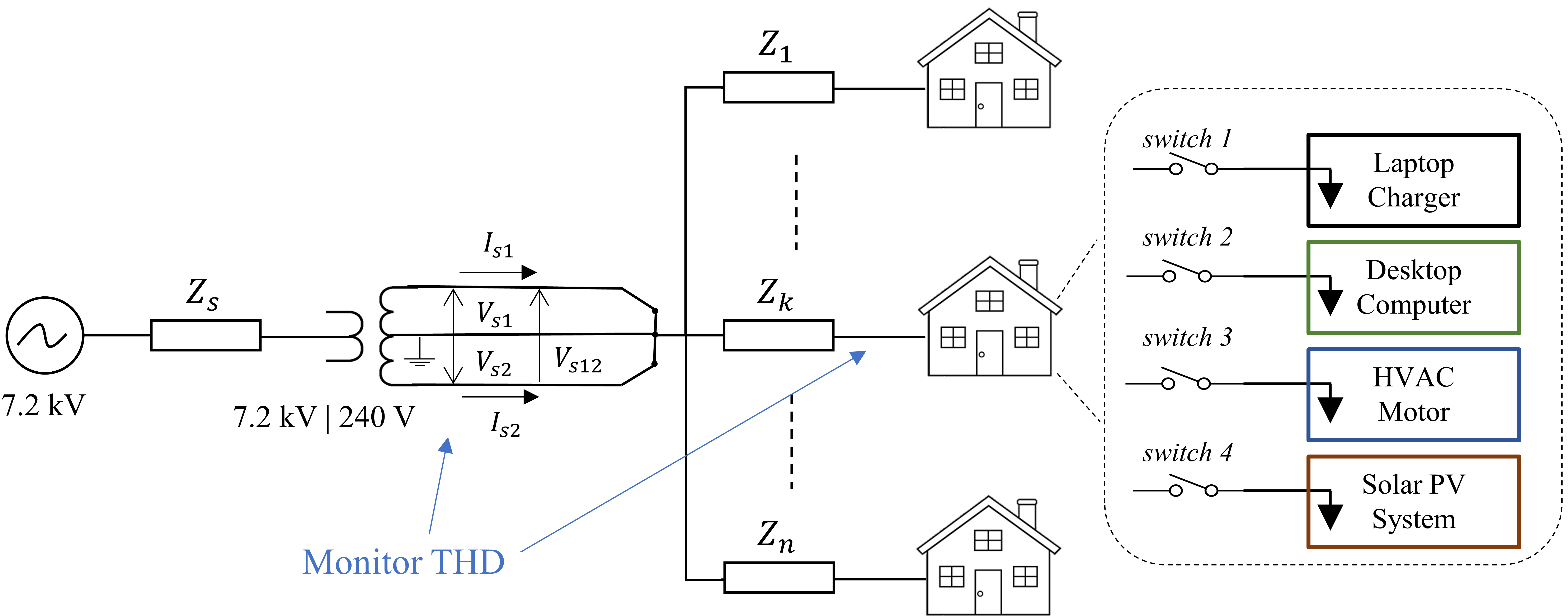}
    \caption{Schematic representation of the simulation setup in PSCAD.}
    \label{fig:house_combo}
\end{figure}

Usually, the residential customers are supplied through a single split-phase connection in the USA. In this work, we are modeling 5 houses that have been connected to a 7.2 kV distribution transformer, as shown in Fig. \ref{fig:house_combo}.
Each home is comprised of four power electronic load combinations shown in Table \ref{tab:load_models} and discussed in detail in \cite{Ankit2022}.

\begin{table}[H]
\renewcommand{\arraystretch}{1.2}
\caption{Power electronics-based load models representing house appliances }
\label{tab:load_models}
\centering
\begin{tabular}{l l}
\hline
 Load model & House appliances \\
\hline
Rectifier + Buck DC-DC converter & Desktop, home entertainment\\
Rectifier + Flyback DC--DC converter & Laptop charger \\
VFD + Induction motor & HVAC, washer, dryer \\
Boost converter + inverter & PV system, EV charger\\
\hline
\end{tabular}
\vspace{-3mm}
\end{table}
\subsection{Data Generation}
The power electronic load combinations were modeled using PSCAD/EMTdc. The steady-state values of current were recorded at the secondary of the distribution transformer as shown in Fig. \ref{fig:house_combo}. A similar process was repeated for several load combinations, and a few scenarios were selected that draw significantly harmonic-rich current  for further analysis as discussed in Section \ref{Results}.

% \subsection{REDD Data}
% The Reference Energy Disaggregation Data (REDD) is a data set that contains low-frequency power consumption data and high-frequency voltage/current waveforms from real
% homes, for the whole house as well as for each individual circuit in
% the house. The data is intended for use in developing disaggregation methods that can predict, from only the whole-home signal, which devices are being used. Data collection procedures and a description of algorithms can be found in \cite{kolter2011redd}. In specific, we utilize high-frequency (16.5 kHz) current/voltage waveform data of the two power mains.
% % (as well as the voltage signal for a single phase). 
% % In order to reduce the data to a manageable size, we have compressed these waveforms using lossy compression. 
% Because the voltage signal in most homes is approximately sinusoidal
% (unlike the current signals, that can vary substantially from a
% sinusoidal wave), zero-crossings of the voltage signal are used to isolate a single cycle of the AC power. For the time spanned by this single cycle, both the current and voltage waveforms are recorded.  However, because the waveforms remain approximately constant for long periods of time, the current and voltage waveforms are only reported at "change points" in the signal. These change points are identified using a method known as total variation
% regularization \cite{kolter2011redd}.

\section{Transformer Degradation Analysis due to Harmonics} \label{sec:trans}
Most power electronic loads are fitted with a single or distributed capacitor at the terminal that helps to maintain a constant dc voltage along with the parasitic inductors. Since there is a periodic change in the load impedance, the current waveform varies from the supplied voltage waveform. The non-sinusoidal current can be represented as a sum of the fundamental and integer multiple of the fundamental \textit{``harmonics"}.

\subsection{Fast Fourier Transform (FFT)} \label{FFT}

One can transform a given sequence in time into its respective frequency components using Discrete Fourier Transform (DFT) \cite{He2016}. FFT is useful for performing the DFT of a sequence. FFT performs the computation of the DFT matrix as a product of sparse factors. The DFT for such a sequence can be given as (\ref{Eqn:DFT}),

\begin{equation}
    X[k] = \sum_{n= 0}^{N-1}x[n]e^{-j2\pi kn /N}
    \label{Eqn:DFT}
\end{equation}

\noindent where \emph{N} is the length of the signal. Since the sampling frequency of the signal is $20kHz$, the maximum represented frequencies are half of the sampling frequency. We try to capture all the representative frequencies in that range. The harmonic components in the measured current are a function of the fundamental $60Hz$ frequency. A frequency scan is performed to identify the magnitude of the current harmonics $I_h$.

Since the measured signal is not an integer multiple, the endpoints of the frequency spectrum are discontinuous. FFT produces a smeared spectral version of the original signal where the energy of one frequency leaks into adjacent frequencies. This phenomenon is known as spectral leakage. To get the best estimate of the current harmonics, we perform a scan of the frequencies adjacent to the integer harmonic frequency. Practices like windowing are utilized to reduce the effect of the non-integer frequencies, but this was not considered as a part of our work.

\subsection{Eddy Current Losses} \label{sec:loss_eddy}

In a residential set-up, most of the losses are due to heating $I^2R$ losses. A residential home consists of both linear and non-linear loads; in this study, we majorly focus on the increase in non-linear residential loads that create harmonics that inadvertently contribute to more losses. The effect is more tremendous at the grid-edge locations where a lot of power electronic loads are connected, for example, the distribution service transformers.

The losses occurring in a transformer can be divided into two categories (\ref{eqn:ploss}), (i) \textit{No-Load losses} and (ii) \textit{Load losses}. In this paper, we study the losses due to non-linear loads. When current flows through a conductor, it generates heat which is either utilized or lost in the surrounding environment.

\begin{equation}
    Tr_{Loss} = Tr_{NL} + Tr_{LL} \label{eqn:ploss}
\end{equation}
The load losses ($Tr_{LL}$) can be further subdivided into a summation of eddy current losses ($Tr_{EC}$) and structural stray losses ($Tr_{ST})$.

The eddy current component can be written as (\ref{eqn:eddy}),
\begin{equation}
    Tr_{EC} = \sum_{h=1}^{h_{max}}I_h^2R_{h} \label{eqn:eddy}
\end{equation} 

\noindent The winding losses increase as a square of the harmonic current component ($I_h$), $R_h$ is the effective resistance comprising of the non-frequency dc component and the resistance that varies with harmonic  content.
When harmonic current flows through the conductive materials of the transformer, it leads to a variation in temperature.

From Newton's law of cooling (\ref{eqn:Newton}),

\begin{equation}
    P\delta t = mc\delta\theta + \alpha \theta A \delta t \label{eqn:Newton}
\end{equation}

\noindent where $P$ is the $I^2R$ losses for material, $m$ is the mass of the material, $c$ is the specific heat capacity of the material, $\delta \theta$ rise of temperature above ambient temperature for time $\delta t$, $A$ is the surface area of the material and $\alpha$ is the emissivity factor.

The change in temperature can be written as (\ref{eqn:temp}) \cite{Tony2012},

\begin{equation}
    \theta (t) = \theta_{final}[1-e^{-t/\tau}] \label{eqn:temp}
\end{equation}

In steady state (\ref{eqn:Newton}), $mc\delta\theta = 0$ and (\ref{eqn:Newton}) can be re-written as (\ref{eqn:new_eqn}),

\begin{equation}
    \begin{array}{l}
        P \delta t = \alpha \theta A \delta t, \\
        \theta = \frac{P}{\alpha A}
    \end{array} \label{eqn:new_eqn}
\end{equation}

\noindent From (\ref{eqn:new_eqn}), we can say that $\theta \propto P \propto I^2R_h$, as more non-linear current passes through the transformer, the ambient temperature of the material changes, resulting in more losses.

\begin{equation}
    R_h = R_{DC}+h^2P_{EC-R}
\end{equation}

\noindent where $R_{DC}$ is the dc-winding resistance at $h^{th}$ harmonic and $P_{EC-R}$ is the winding eddy current loss factor, that ranges between 0.01 in low voltage transformers to 0.10 for substation transformers. For our study, we consider $P_{EC-R}$ as 0.05.

By replacing $R_h$ in (\ref{eqn:eddy}), we get

\begin{equation}
    Tr_{EC} = I_1^2R_{DC}+\sum_{h=3,5,7,...}^{h_{max}}I_h^2h^2P_{EC-R} \label{eqn:eddy_loss}
\end{equation}

\noindent The first term of (\ref{eqn:eddy_loss}) $I_1^2R_{DC}$ is the non-frequency dependant part, and $I_h^2h^2P_{EC-R}$ reflects the frequency dependant part of the transformer eddy current losses. Thus, the harmonic driven transformer eddy current losses can be summarized as (\ref{eqn:eddy_final}),

\begin{equation}
    Tr_{EC} = P_{EC-R}\sum_{h=1}^{h_{max}}I^2h^2 \label{eqn:eddy_final}
\end{equation}
\subsection{Harmonic Loss Factor \& Transformer Derating} \label{k-factor}

Harmonic loss factor ($F_{HL}$) is defined as the ratio of the total loss due to eddy current due to harmonics and the winding current losses in the absence of harmonics \cite{IEEEC57}. It is expressed as (\ref{eqn:f-hl}),

\begin{equation}
    F_{HL} = \frac{\sum_{h=1}^{h_{max}}I_h^2 h^2}{\sum_{h=1}^{h_{max}}I_h^2}
    \label{eqn:f-hl}
\end{equation}

 To reduce the loss-of-life of a transformer due to an increased non-linear current, they are usually derated (i.e., reduced transformer loading). The derating \% helps to understand the transformer operational capability below its rating to prolong its duration.

\section{Results and Discussion} \label{Results}
\subsection{Simulation Setup}
\begin{figure}
    \centering
    \includegraphics[width=1\columnwidth]{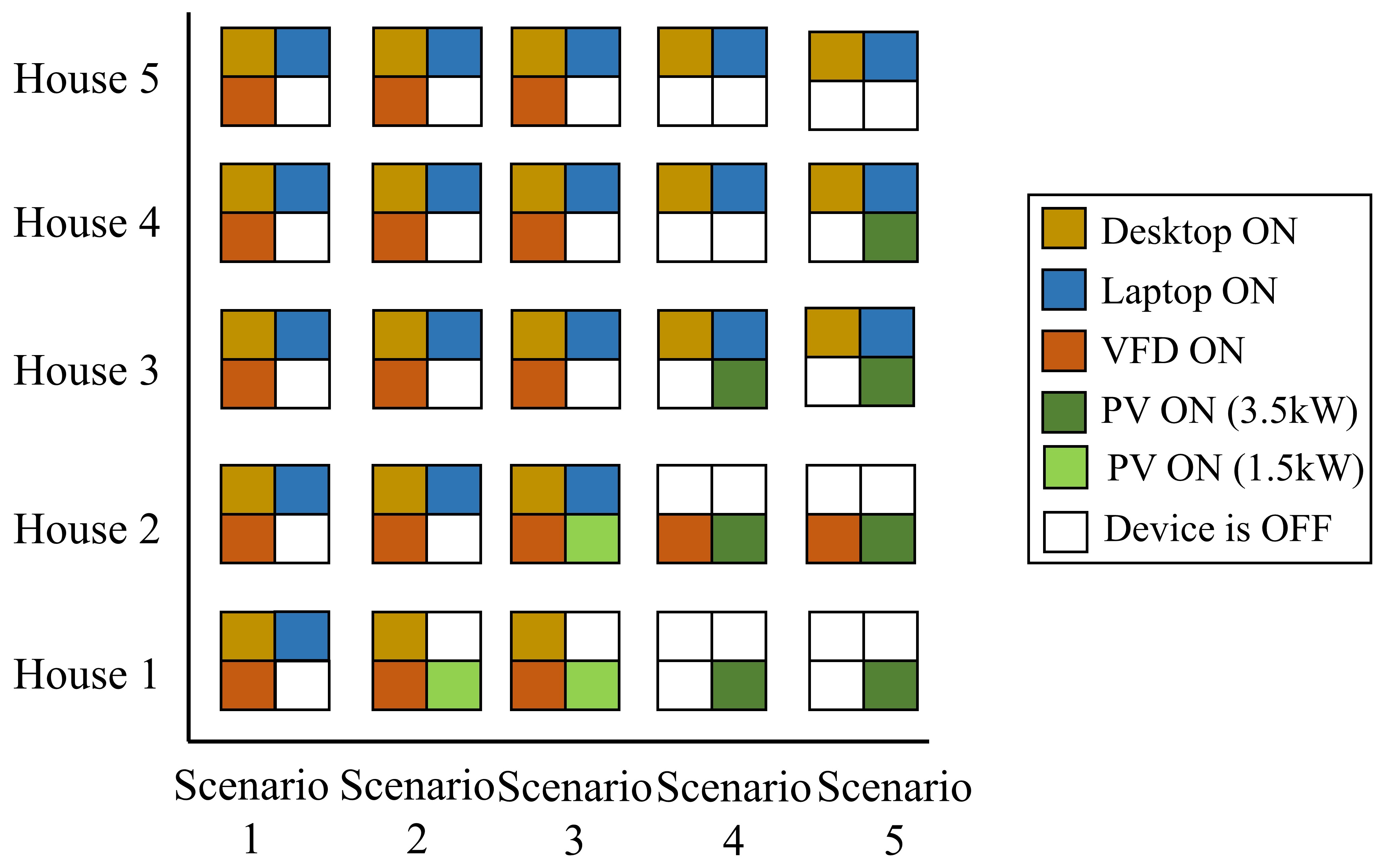}
    \caption{Load combination for 5 houses corresponding to a different scenario, each scenario represents the peak load condition with a given PV penetration.}
    \label{fig:cases}
\end{figure}

To understand the effect of different loading scenarios on transformers,  the simulation setup described in Fig. \ref{fig:house_combo} is used. A total of 5 scenarios are constructed with different power electronic load combinations to analyze the impact on the transformer, as shown in Table \ref{tab:scenario}. All these 5 scenarios are assumed to represent the peak loading condition for a given transformer with increasing solar PV units. Scenarios 1, 2, and 3 are evening peaking cases where solar generation. Whereas scenarios 4 and 5 have solar generation high enough to cause a reverse power peak during daytime when the load is low. It is assumed that 1 PV unit generates 3.5 kW and 1.5 kW during the daytime and evening, respectively. The exact loading condition (mix of VFD, laptop, desktop, and PV) of each of the 5 houses connected to a service transformer in each scenario  is shown in Fig. \ref{fig:cases}. 
% The currents were recorded for all the houses and at the secondary of the service transformer. 
% From our analysis and simulations we were able to determine the mix of VFD, laptop, desktop and PV load combinations that would contribute the highest harmonic distortion, Fig. \ref{fig:cases} shows the various combinations of VFD and PV loads that were used for the analysis.

\begin{table}[]
\renewcommand{\arraystretch}{1.1}
\caption{Scenarios of different load combinations with increasing PV penetration}
\label{tab:scenario}
\begin{tabular}{cccccc}
\hline
\multirow{2}{*}{Scenarios} & \multirow{2}{*}{PV units} & Peak Load & Total PV & Total other & \multirow{2}{*}{Net load} \\
 &  & Time & generation & load &  \\ \hline
1 & 0 & evening & 0 kW & 9.5 kW & 9.5 kW \\
2 & 1 & evening & 1.5 kW & 9.5 kW & 8 kW \\
3 & 2 & evening & 3 kW & 9.5 kW & 6.5 kW \\
4 & 3 & day & 10.5 kW & 2.5 kW & -8 kW \\
5 & 4 & day & 14 kW & 2.5 kW & -11.5 kW \\ \hline
\end{tabular}
\end{table}

\begin{figure}
    \centering
    \includegraphics[width=0.48\textwidth]{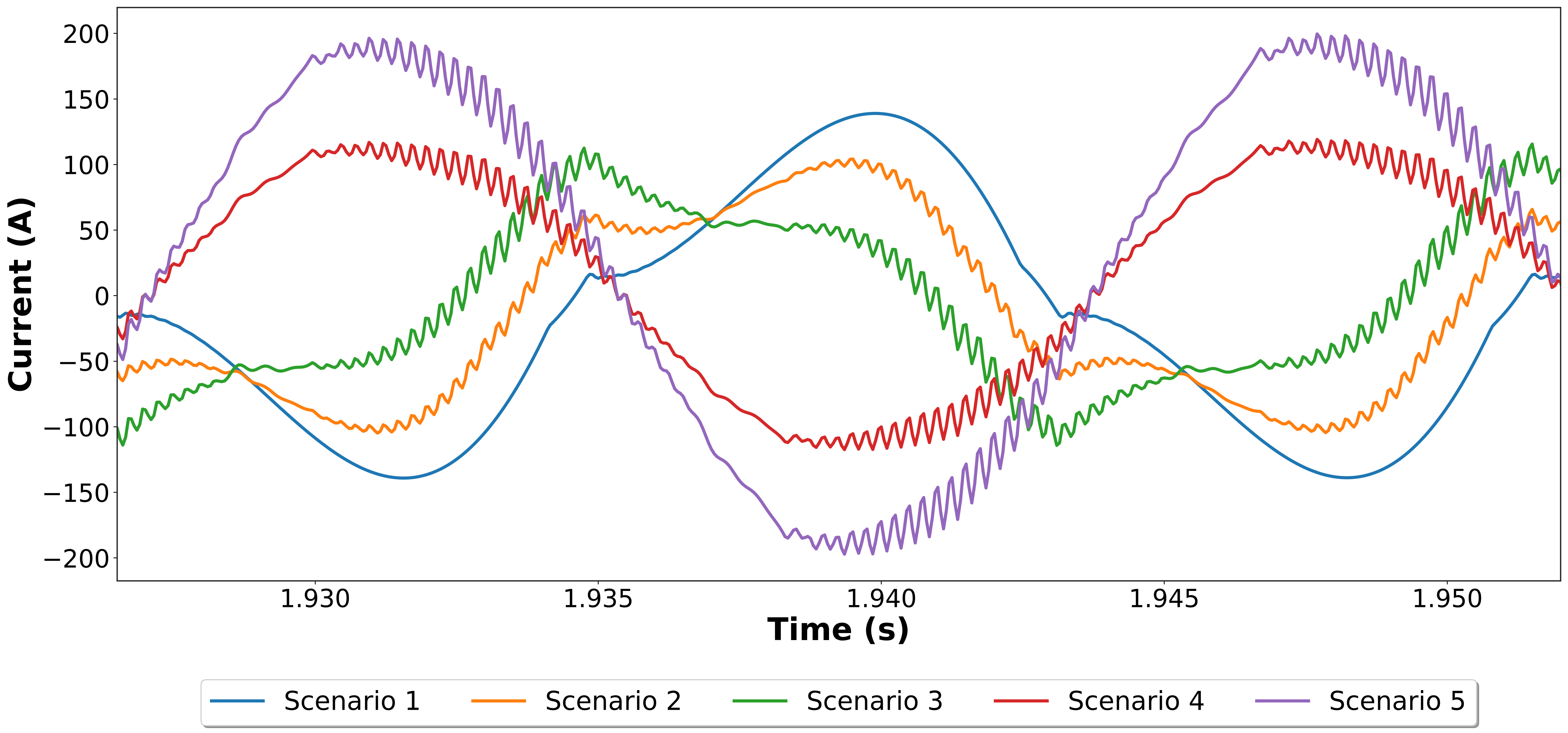}
    \caption{Equivalent transformer secondary current for different scenarios.}
    \label{fig:trans_curr}
\end{figure}

\subsection{Impact of increasing PV on Transformer Current Harmonics and Eddy Losses}
\begin{figure}
    \centering
    \includegraphics[width=0.48\textwidth]{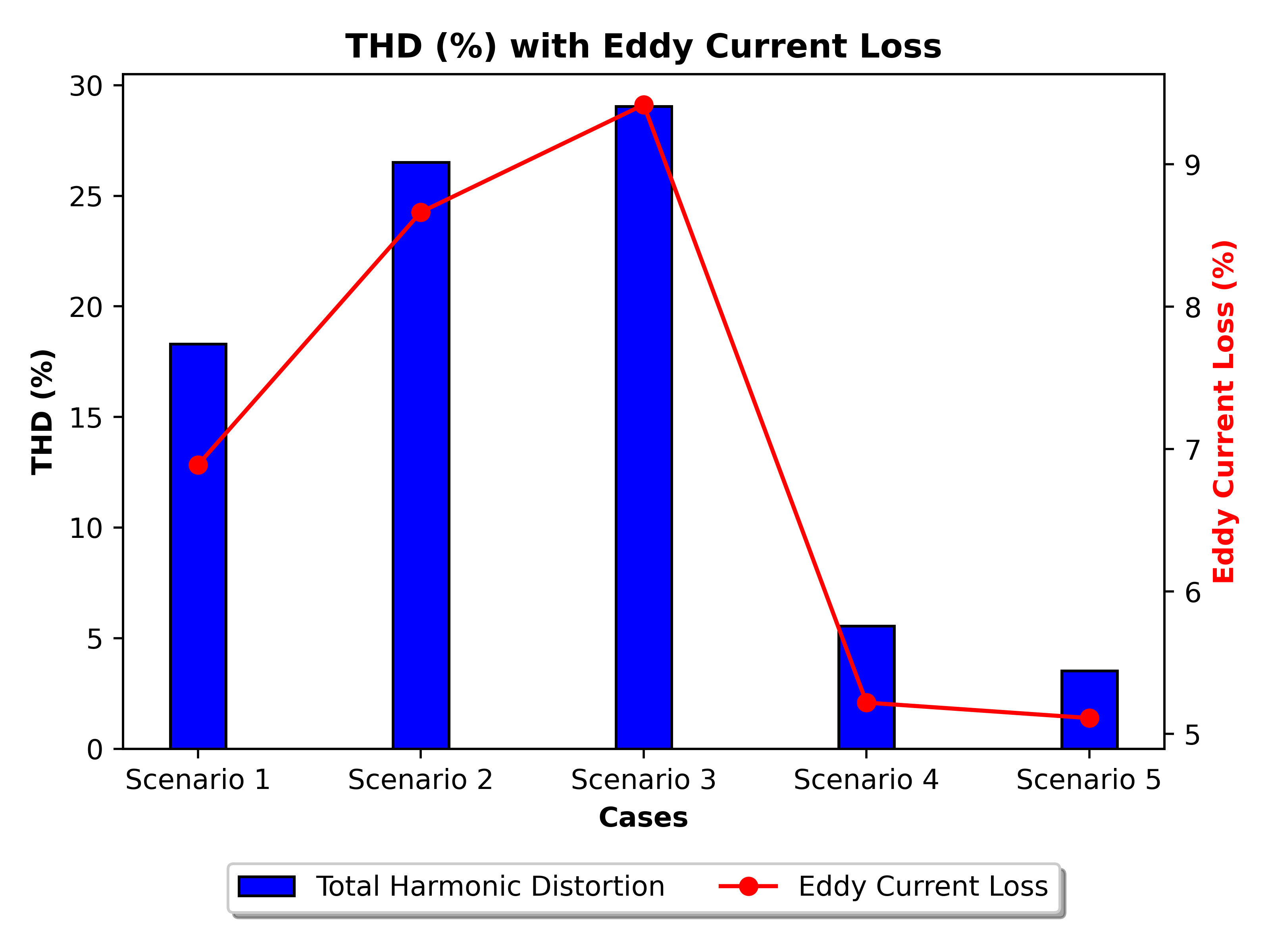}
    \caption{Variation of THD(\%) along with eddy current losses for different load combinations and PV penetration.}
    \label{fig:eddy_loss}
\end{figure}

\begin{figure}
    \centering
    \includegraphics[width=0.48\textwidth]{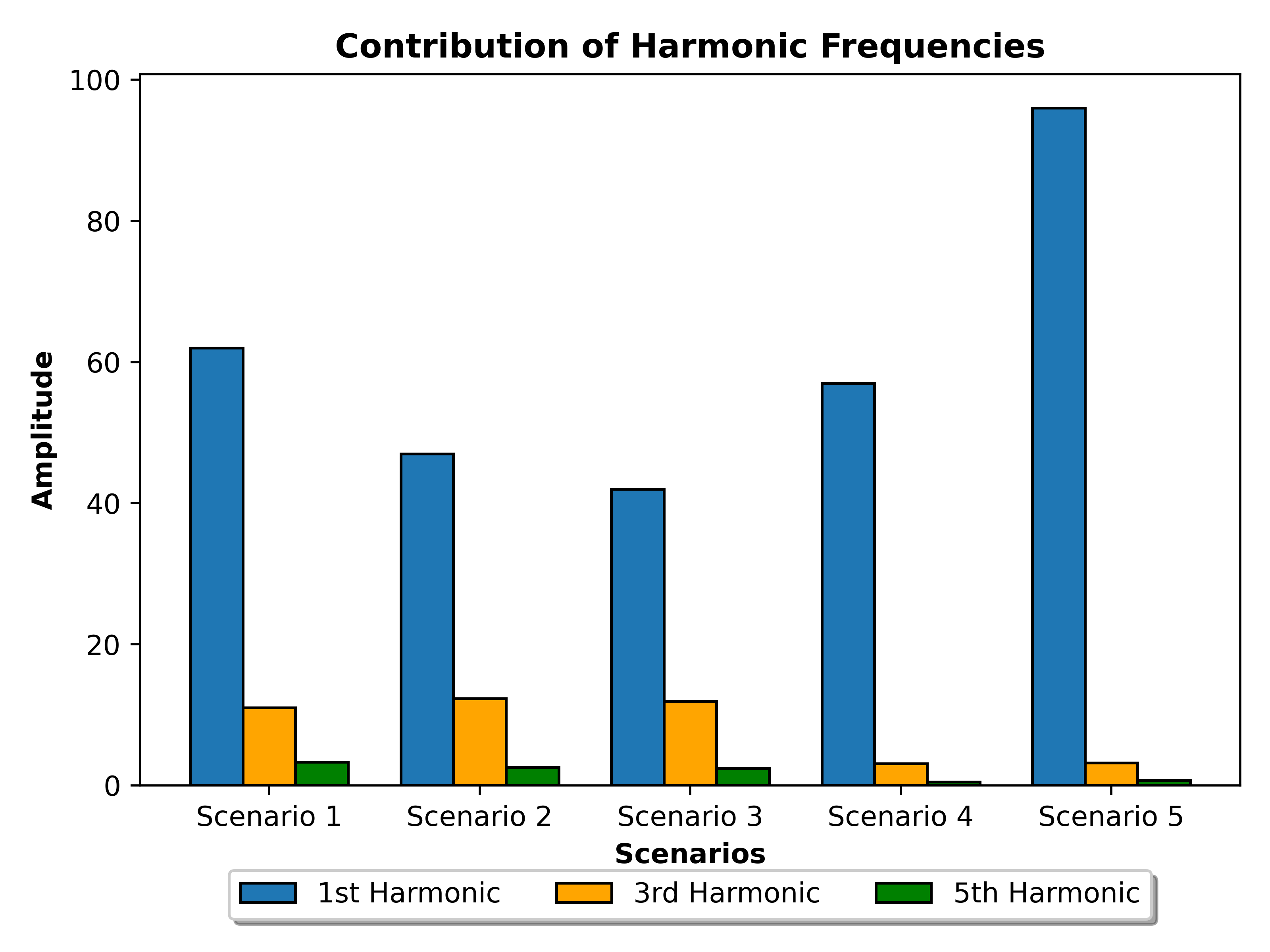}
    \caption{The harmonic frequency spectrum of the secondary transformer under various load combinations and PV penetrations.}
    \label{fig:freq}
\end{figure}

\begin{figure}
    \centering
    \includegraphics[width=0.48\textwidth]{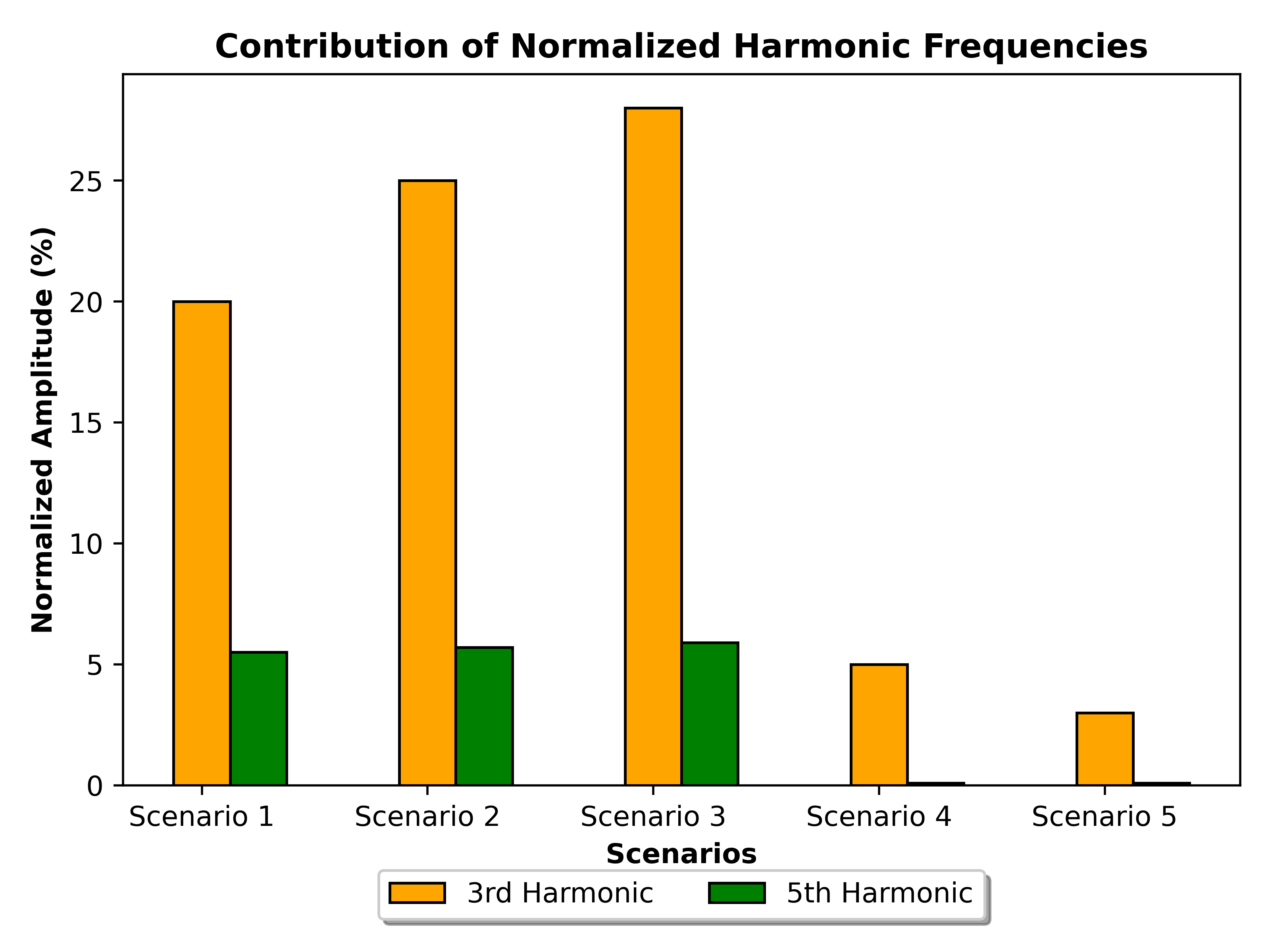}
    \caption{Variation of normalized frequency spectrum of $3^{rd}$ \& $5^{th}$ harmonic with respect to fundamental current magnitude with different load and PV penetration scenarios.}
    \label{fig:freq_norm}
\end{figure}

The current waveforms drawn by the transformer secondary in all 5 scenarios are shown in Fig. \ref{fig:trans_curr}. The reverse current can be noticed in scenarios 3 and 4 due to high solar PV generation. These waveforms were analyzed to understand the harmonic contribution in each scenario. The harmonic contents of the measured current signals were extracted using FFT as described in Section  \ref{FFT}. The THD(\%) for each of the scenarios was calculated using (\ref{eqn:thd}),
\begin{equation}
    THD = \frac{\sqrt{\sum_{h=2}^{h_{max}}I_h^2}}{I_1} \label{eqn:thd}
\end{equation}
Where $I_h$ denotes the $h^{th}$ harmonic current magnitude. Eddy current losses calculations are based on the discussion in Section \ref{sec:loss_eddy}. THD(\%) and corresponding eddy losses for each scenario are shown in Fig. \ref{fig:eddy_loss}. It is observed in scenarios 1, 2, and 3 that increasing solar PV units cause more harmonic distortion in transformer currents. On the contrary, in scenarios 4 and 5, it is observed that the addition of PV generation decreases THD. To understand this pattern, we need to look at the frequency spread of $1^{st}$, $3^{rd}$, and $5^{th}$ harmonic of the current as shown in Fig. \ref{fig:freq}. Observing scenarios 1,  2, and 3, we find that the increasing PV generation reduces the net fundamental current magnitude as expected. However, the $3^{rd}$ harmonic current magnitude remains the same in scenarios 1, 2, and 3, as shown in Fig. \ref{fig:freq}. It can be inferred that the PV inverter is primarily compensating for the fundamental component of the load currents, not the $3^{rd}$ harmonic. This leads to an increased \% of $3^{rd}$ harmonic in scenario 3, as shown in Fig. \ref{fig:freq_norm}, resulting in high THD(\%). 

On the other hand, scenarios 4 and 5 have very high PV generation but a much lower amount of other power electronics load compared to scenarios 1-3. Therefore, the net current is mainly composed of PV current. Since PV units are mandated to maintain less than 5\% THD, they come equipped with harmonic filters by the vendors, as discussed in \cite{Ankit2022}. Therefore, in scenarios 4 and 5, harmonic distortions are the lowest.

Transformer eddy losses tend to follow the current THD \% and are shown in Fig. \ref{fig:eddy_loss}. It can be seen that scenario 3 has the highest eddy losses, close to 30\%.

\subsection{Impact on Transformer Derating}
All 5 scenarios represent the peak loading situation for the given PV penetration. It is important for the transformer derating analysis as it is usually performed in a full-loading condition. Note that the up to 2 PV unit penetration (scenario 1-3) peak loading is assumed to occur during the evening when other loads are high. Whereas, for higher penetration (scenarios 4-5), solar PV generation can create its own reverse power peak during the daytime.

The harmonic loss factor helps us to understand the \% at which the transformers should be operated to prolong its life. The derating \% for the different scenarios is shown in the last column of the Table \ref{tab:scenario}. The worst derating is observed in scenario 3, where the transformer operates at 75.88\% of its rated capacity resulting in significant loss of life. If the penetration of power electronic loads are further increased the transformers would need to be further derated for their operation.

\begin{table}[]
\centering
\caption {The overall impact of increasing PV penetration on transformer degradation in terms of THD, eddy current losses, and transformer derating}
\label{tab:results}
\begin{tabular}{ccccc}
\hline
Scenarios & THD (\%) & $Tr_{EC}$ (\%)  & Derating (\%) \\ \hline
1 & 18.30 & 6.89  & 85.59 \\
2 & 26.51 & 8.66  & 78.59 \\
3 & 29.05 & 9.41  & 75.88 \\
4 & 5.55 & 5.22 & 98.01 \\
5 & 3.52 & 5.11  & 98.95 \\ \hline
\end{tabular}
\vspace{-4mm}
\end{table}

Total impact on the transformer in terms of THD(\%), eddy losses, and derating are listed in Table \ref{tab:results}. Overall, if PV units come equipped with a filter as mandated by the standards, their individual effect on the transformer loading is positive. Therefore, in the presence of low power-electronic loads (scenarios 4 \& 5), increasing PV penetration has a positive impact on transformer degradation. However, in the presence of high power-electronic loads (scenarios 1-3), increasing PV penetration may have adverse impacts on transformer degradation since it is not able to compensate for the $3^{rd}$ harmonic consumed by the loads such as VFDs, thus increasing $3^{rd}$ harmonic contribution.

\section{Conclusion} \label{Conclusion}

The proposed work represents the challenges faced by a low-voltage distribution transformer due to the high penetration of a power electronic-dominated residential infrastructure via EMTP simulations. By operating principle, transformers are linear devices, and the addition of non-linear power electronic loads makes their operation non-linear. The harmonic load currents increase the losses in a transformer. The non-linear current causes a rise in temperature that affects the effective resistance of the transformer, as discussed in the Section \ref{sec:trans}. The eddy current losses depend on the magnitude of the harmonic current, and higher THD(\%) contributes to more eddy current losses. Integration of PV resources to compensate for the transformer loading had an adverse effect with increased levels of $3^{rd}$ harmonic. Such scenario's are particularly visible for scenario 2 \& 3 where the load of the network is mostly compensated by PV generation. Although, under low power electronic loading conditions, the addition of PV helped to reduce THD(\%) in the transformer.

To have a better estimate of the transformer performance, detailed power electronic models are necessary. However, the scalability of inverter models in different simulation tools beyond a certain number is infeasible. Thus developing mathematical models of harmonic loads via frequency coupled matrix (FCM) will be considered in future work to generate high-fidelity time-series data for further analysis. 

\bibliographystyle{IEEEtran}
\bibliography{HELM_2023.bib}
\end{document}